\documentclass[]{emulateapj}

\begin{document}

\newcommand{\kms}{\hbox{km s$^{-1}$}}	\newcommand{\msun}{$M_{\odot}$} 
\newcommand{\mas}{mas yr$^{-1}$}	\newcommand{\vsini}{$v\sin{i}$} 
\newcommand{\teff}{$T_{\rm eff}$}	\newcommand{\logg}{$\log{g}$} 

\slugcomment{Submitted to ApJ}

\title{ Proper Motions and Trajectories for 16 Extreme Runaway and 
	Hypervelocity Stars\altaffilmark{1} }

\altaffiltext{1}{Based on observations with the NASA/ESA {\it Hubble Space 
Telescope\/} obtained at the Space Telescope Science Institute, and from the data 
archive at STScI, which is operated by the Association of Universities for Research 
in Astronomy, Inc., under NASA contract NAS5-26555. }

\author{Warren R.\ Brown$^2$, Jay Anderson$^3$, Oleg Y.\ Gnedin$^4$, 
	Howard E.\ Bond$^{3,5}$, Margaret J.\ Geller$^2$, and Scott J.\ Kenyon$^2$}

\affil{	$^2$Smithsonian Astrophysical Observatory, 60 Garden St, Cambridge, MA 02138\\
	$^3$Space Telescope Science Institute, 3700 San Martin Dr., Baltimore, MD 21218\\
	$^4$Department of Astronomy, University of Michigan, Ann Arbor, MI 48109\\
	$^5$Department of Astronomy \& Astrophysics, Pennsylvania State University, University Park, PA 16802
	}

\noindent \email{wbrown@cfa.harvard.edu;jayander@stsci.edu;ognedin@umich.edu }

\shorttitle{ Trajectories for 16 Runaways and HVSs }

\shortauthors{Brown, Anderson, Gnedin, et al.}

\begin{abstract}

	We measure proper motions with the {\it Hubble Space Telescope} for 16 
extreme radial velocity stars, mostly unbound B stars in the Milky Way halo.  
Twelve of these stars have proper motions statistically consistent with zero, and 
thus have radial trajectories statistically consistent with a Galactic center 
``hypervelocity star'' origin.  The trajectory of HE 0437--5439 is consistent with 
both Milky Way and Large Magellanic Cloud origins.  A Galactic center origin is 
excluded at 3$\sigma$ confidence for two of the lowest radial velocity stars in our 
sample, however.  These two stars are probable disk runaways and provide evidence 
for $\sim$500 \kms\ ejections from the disk.  We also measure a significant proper 
motion for the unbound sdO star US~708.  Its 1,000 \kms\ motion is in some tension 
with proposed supernova ejection models, but can be explained if US~708 was ejected 
from the stellar halo.  In the future, we expect {\it Gaia} will better constrain 
the origin of these remarkable unbound stars.

\end{abstract}


\section{INTRODUCTION}

	The Milky Way is a gravitationally bound system of a few $10^{11}$ stars,
but it also hosts some unbound stars.  Where these unbound stars come from is an
open question.  It is very difficult to explain unbound main-sequence stars by
supernova explosions in close binary systems \citep{blaauw61} or by dynamical
encounters between binaries \citep{poveda67}, the mechanisms that explain disk
``runaway'' stars.  Simulations show that stars cannot be launched at speeds greater
than their binary orbital velocity, and this velocity cannot exceed the escape
velocity from a star's surface, since that would require the stars to be orbiting
inside one another \citep{leonard91}.  The escape velocity from the surface of
main-sequence stars is comparable to the 500 -- 600 \kms\ local Galactic escape
velocity \citep{piffl14}.  Only dynamical interactions with an object more massive
and compact than a star can easily explain unbound main-sequence stars.  
\citet{hills88} predicted that three-body exchange interactions between stars and
the central massive black hole (MBH) will inevitably unbind stars from the Galaxy,
and he called such objects ``hypervelocity stars'' (HVSs).  We therefore refer to
unbound main-sequence stars in this paper as HVSs.

	The first HVS was discovered by \citet{brown05}, and a couple dozen B-type 
HVSs are now known in the Milky Way halo \citep{edelmann05, brown06, brown06b, 
brown07a, brown07b, heber08, brown09a, brown12b, brown14, zheng14}.  These stars 
have radial velocities up to $+700$ \kms; no star moving towards us with a 
comparable radial velocity has ever been observed.  To date, detailed spectroscopic 
analyses of the B-type HVSs find that they are main-sequence B stars at 10 -- 100 
kpc distances \citep{edelmann05, przybilla08, przybilla08b, lopezmorales08, 
brown12c, brown13b}.  The short lifetimes of B stars require that these unbound 
stars were ejected from a region with recent star formation, such as the Galactic 
disk or the Galactic center.

	Full three-dimensional trajectories for the HVSs are a crucial test of their 
origin.  Measuring the trajectories for the HVSs, however, requires accurate 
distances and absolute proper motions.  Distance estimates to individual stars are 
determined by comparing either spectroscopic \teff\ and \logg\ or broadband colors 
to stellar evolution tracks.  Typical precision is 15\% \citep{brown14}; the 
accuracy depends on the choice of stellar evolution tracks (i.e.\ assumptions about 
metallicity).  Proper motions are equally difficult to measure, because HVSs are 
distant and they should be on radial trajectories.  Expected HVS proper motions are 
typically $\lesssim1$ \mas\ and thus cannot be measured with ground based surveys.  
The only unbound star with a statistically significant proper motion measurement to 
date is the runaway B star \hbox{HD 271791}, a star that was ejected in the 
direction of Galactic rotation from the outer disk \citep{heber08}.  \hbox{HIP 
60350} is another example of a high velocity B star with a significant proper motion 
measurement, but it is unbound at less than $1\sigma$ significance 
\citep{irrgang10}.  Here, we present {\it Hubble Space Telescope (HST)} proper 
motion measurements for a sample of 16 stars with extreme radial velocities, 12 of 
which are unbound to the Milky Way in radial velocity alone.

	Proper motions are important because they can discriminate between Galactic 
center and Galactic disk origins.  The star \hbox{US 708} (hereafter \hbox{HVS 2}), 
for example, is a helium-rich sdO star with a +708 \kms\ heliocentric radial 
velocity \citep{hirsch05}.  sdO stars are the remnants of low mass stars; helium 
rich sdO stars are possibly the merger product of two helium white dwarfs 
\citep{heber09}.  If \hbox{HVS 2} were ejected by the central MBH, its velocity 
vector will point away from the Galactic center.  If \hbox{HVS 2} were ejected by a 
Type Ia supernova explosion \citep{justham09, wang09}, its velocity vector can point 
from anywhere in the Milky Way.

%via a triple star disruption and the subsequent merger of a pair of white dwarfs 
%\citep{perets09a}

	Generally speaking, lower velocity stars are more likely to be disk 
runaways.  The maximum ejection velocity in the \citet{blaauw61} supernova mechanism 
is the sum of the supernova kick velocity and the orbital velocity of the progenitor 
binary, or about 250 \kms\ for an ejected main-sequence B star \citep{portegies00}.  
Velocities up to 500 \kms\ may be possible for B stars ejected by extremely 
asymmetric core-collapse supernovae in contact binaries \citep{tauris15}. The 
maximum ejection velocity in the dynamical mechanism depends on the most massive 
star in the encounter, and can also reach 500 \kms\ in extreme scenarios 
\citep{leonard91, gvaramadze09}.  Simulations suggest that 99\% of dynamical 
ejections occur at velocities $<$200 \kms\ \citep{perets12}.  We test for the 
existence of $\sim$500 \kms\ runaway ejections by including four bound velocity 
outliers in our sample.

	The star HE 0437--5439 (hereafter \hbox{HVS 3}), an unbound 9 \msun\ 
main-sequence B star located 16 deg from the Large Magellanic Cloud (LMC), presents 
another puzzle \citep{edelmann05}.  The lifetime of this star is many times shorter 
than its flight time from the Milky Way, suggesting an LMC origin.  An LMC origin 
requires a 1,000 \kms\ ejection, however, and thus an unseen MBH in the LMC 
\citep{gualandris07}.  A Milky Way origin, on the other hand, requires that 
\hbox{HVS 3} be a blue straggler to account for its lifetime.  In other words, the 
progenitor must have been a binary system ejected at \hbox{$>800$} \kms\ that 
subsequently merged as it traveled away from the Milky Way.  A binary MBH could 
eject stellar binaries as HVSs \citep{lu07}, or else a single MBH could eject 
binaries by triple disruption \citep{perets09a}.  All of these models have very low 
ejection rates.  The LMC and Galactic center origins for \hbox{HVS 3} differ by 1.5 
\mas\ in proper motion.  Using two epochs of {\it HST} imaging, \citet{brown10b} 
found that \hbox{HVS 3} is moving away from the Milky Way.  \citet{irrgang13} argue 
that both Milky Way and LMC origins, given the systematic uncertainties, are 
consistent with the measurements.  A third epoch of imaging is needed.

	In principle, {\it HST} proper-motion measurements have sufficient 
precision to determine the origin of our unbound stars.  Using the best data 
reduction and measurement techniques, it is possible to achieve 0.01 pixel 
astrometric precision on well-exposed stars \citep[e.g.][]{bellini14}.  The Advanced 
Camera for Surveys (ACS) Wide-Field Channel has 50 mas pixels, and the Wide Field 
Camera 3 (WFC3) Ultraviolet-Visible (UVIS) Channel has 40 mas pixels.  
For a pair of measurements separated by a 3 year time baseline, we can thus expect 
0.2 \mas\ precision.  We must rely on background galaxies to define our absolute 
reference frame, however, and both the number and spatial distribution of useful 
background galaxies introduce a systematic uncertainty to the measurements. In some 
cases, a sequence of short and long exposures is used to link faint background 
galaxies to bright HVSs, adding additional uncertainty.

	For the sake of clarity, we quote proper motion uncertainties that are the 
sum in quadrature of the statistical (stellar) and systematic (background galaxy) 
proper motion uncertainties throughout this paper.  Our average total uncertainty is 
$\pm0.8$ \mas, a 6-fold improvement over existing proper motion measurements.

	In Section 2 we define the sample of 12 unbound and four bound stars, and 
present new spectroscopy and stellar atmosphere fits for the four bound stars.  In 
Section 3 we describe the {\it HST} imaging, image reduction, and proper motion 
measurements.  Twelve stars have proper motions statistically consistent with zero, 
and thus largely radial trajectories.  In Section 4 we evaluate the constraints on 
a Galactic center origin, accounting simultaneously for radial velocity, distance, 
and proper motion errors.  Thirteen stars have trajectories consistent with a 
Galactic center origin given the measurement errors, while three stars are 
inconsistent with a Galactic center origin at $>$3$\sigma$ confidence. In Section 5 
we discuss the objects with significant proper motions, the runaway stars \hbox{HVS 
2}, B~711, and B~733.  We also update our constraints on HVS 3, consistent with 
either a LMC or Milky Way origin.  We conclude in Section 6.

\section{THE SAMPLE}

	We select our sample of 16 stars for their extreme radial velocities.  We 
targeted all 12 unbound HVSs known prior to 2008 March, plus four bound velocity 
outliers from the HVS Survey of \citet{brown07b}.  Table \ref{tab:hvs} lists the 
coordinates and other observed properties of the stars.  We refer to the 12 unbound 
stars as \hbox{HVS 1} \citep{brown05}, \hbox{US 708} = \hbox{HVS 2} 
\citep{hirsch05}, HE $0437-5439$ = \hbox{HVS 3} \citep{edelmann05}, \hbox{HVS 4} - 
\hbox{HVS 5} \citep{brown06}, \hbox{HVS 6} - \hbox{HVS 7} \citep{brown06b}, 
\hbox{HVS 8} - \hbox{HVS 10} \citep{brown07b}, and \hbox{HVS 12} - \hbox{HVS 13} 
\citep{brown09a}.  We refer to the four bound velocity outliers as B~434, B~485, 
B~711, and B~733, corresponding to their target numbers in the HVS Survey 
\citep{brown07b}.

	Detailed analyses of the unbound stars indicate that, except for \hbox{HVS 
2}, they are main-sequence B stars at 50 - 100 kpc distances.  This conclusion is 
based on stellar atmosphere fits to high resolution echelle spectra of \hbox{HVS 3} 
\citep{przybilla08, bonanos08}, \hbox{HVS 5} \citep{brown12c}, \hbox{HVS 7} 
\citep{przybilla08b}, and \hbox{HVS 8} \citep{lopezmorales08}, and fits to moderate 
resolution, high signal-to-noise (S/N) spectra of the other HVSs \citep{brown14}.  
Here we describe spectroscopy of the four bound velocity outliers.

\begin{figure}		% FIGURE 1:  Spectra
 \plotone{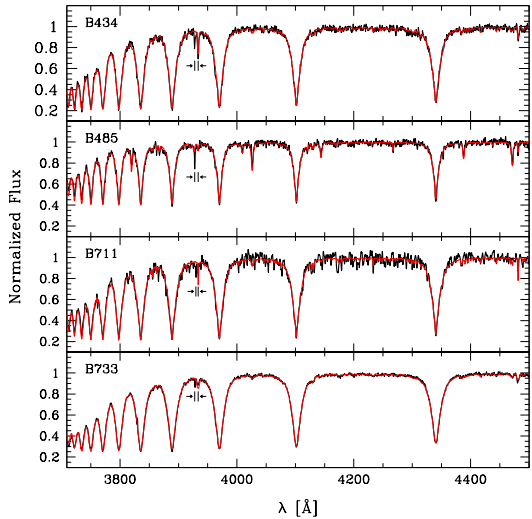}
 \caption{ \label{fig:spec}
	Spectra of the bound velocity outliers, continuum-normalized and shifted to 
rest-frame (in black), compared to their best-fitting stellar atmosphere model (in 
red).  The wavelength separation between the pair of Ca {\sc ii} $\lambda$3933 lines 
(marked with arrows), the lefthand one due to local interstellar medium absorption, 
visibly shows each star's large radial velocity.  We use the hydrogen Balmer lines 
to measure \teff, \logg, and \vsini.}
 \end{figure}

\subsection{New Spectroscopy}

	We obtained 6.5m MMT Blue Channel spectroscopy of the four bound velocity
outliers in four different observing runs in 2006 February, 2008 February, 2010
March, and 2014 March.  We used the 832 l mm$^{-1}$ grating in 2nd order with a 1
arcsec slit, providing 1.0 \AA\ spectral resolution over the range 3600 - 4500 \AA.  
In addition, we observed B~485 a second time with a 0.75 arcsec slit to test higher
0.75 \AA\ spectral resolution.  We paired every observation with a He-Ne-Ar
comparison lamp exposure for wavelength calibration.  We chose exposure times to
achieve S/N=50 - 100 per resolution element in the continuum.

	We process and extract the 1D spectra using IRAF.\footnote{IRAF is 
distributed by the National Optical Astronomy Observatories, which are operated by 
the Association of Universities for Research in Astronomy, Inc., under cooperative 
agreement with the National Science Foundation.} We measure radial velocities using 
the cross-correlation package RVSAO \citep{kurtz98}.  We then measure effective 
temperature \teff, surface gravity \logg, and projected rotation \vsini\ using 
stellar atmosphere models as described in \citet{brown14}.  Figure \ref{fig:spec} 
compares the best-fit stellar atmosphere models to the observed spectra.  All four 
bound velocity outliers are late B-type stars, and B~733 has significant projected 
rotational velocity \vsini=$240\pm30$ \kms.

\begin{figure}		% FIGURE 2:  Teff
 \plotone{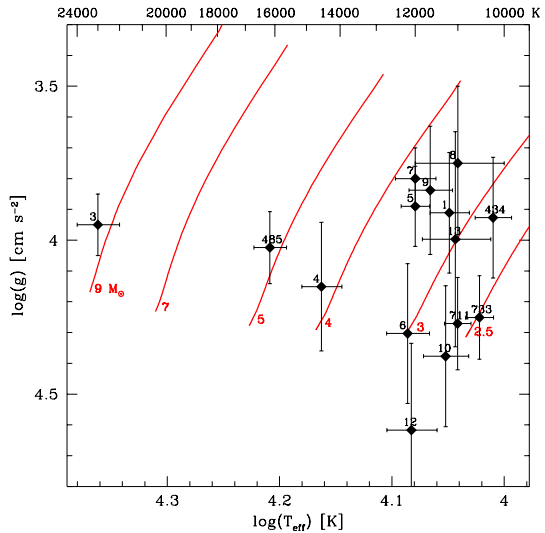}
 \caption{ \label{fig:teff}
	Effective temperature \teff\ and surface gravity \logg\ compared with Padova 
solar metallicity main-sequence tracks (red lines) for our sample of stars.  
Stars are labeled by their HVS/B identifier number. }
 \end{figure}

\subsection{Stellar Parameters}

	Our sample contains 15 B-type stars plus the sdO star \hbox{HVS 2}.  Figure 
\ref{fig:teff} displays the adopted effective temperature \teff\ and surface gravity 
\logg\ for the 15 B-type stars in our sample, plotted in comparison to Padova solar 
metallicity main-sequence tracks \citep{girardi04, marigo08, bressan12}.  \hbox{HVS 
2} does not appear because it is a \teff=45,560 K sdO star.  The clumping of stars 
in Figure \ref{fig:teff} reflects the HVS Survey target selection of stars with the 
colors of 3 \msun\ main sequence stars.

	Notably, all four stars with high resolution echelle spectroscopy are fast
rotators.  Another five B-type stars have statistically significant \vsini\
$\ge150$ \kms\ on the basis of our moderate resolution MMT spectroscopy.  Fast
rotation is the unambiguous signature of a main-sequence B star.  For reference, the
mean \vsini\ of late B-type stars is 150 \kms\ \citep{abt02, huang06a}.  Evolved
horizontal branch stars of the same temperature and surface gravity, on the other
hand, are slow rotators with \vsini\ $<7$ \kms\ \citep{behr03b}.  Thus these B-type
stars are main-sequence stars.

	We compare the measured stellar atmosphere parameters to Padova 
main-sequence stellar evolution tracks to estimate luminosities \citep{girardi04, 
marigo08, bressan12}.  Propagating the measurement uncertainties through the stellar 
evolution tracks implies that our luminosity estimates are precise to 30\% in 
luminosity, or 15\% in distance.  The precision is relatively poor because surface 
gravity is our primary constraint on evolutionary status, and the luminosities of B 
stars increase with age.

	The choice of stellar evolution tracks is a source of systematic 
uncertainty.  With the exception of \hbox{HVS 3} and \hbox{HVS 7}, which have solar 
iron abundance \citep{przybilla08, przybilla08b}, the metallicity of the HVSs is 
unconstrained.  If the HVSs are systematically metal-rich or metal-poor, our 
distance estimates could be systematically wrong by $\sim$25\% 
\citep[e.g.][]{bressan12}.  Given that the HVSs are relatively short-lived 
main-sequence B stars, however, we consider solar metallicity a reasonable assumption.

	We calculate heliocentric distance using the de-reddened $g$-band point 
spread function apparent magnitudes from Sloan Digital Sky Survey (SDSS) Data 
Release 10 (DR10) \citep{ahn14}, except for \hbox{HVS 3} which has a $V$-band 
measurement \citep{bonanos08}.  The SDSS magnitudes have a typical precision of 2\%;
however in some cases the DR10 $g$-band magnitude differs from previous SDSS data 
release values by as much as 10\%.  The SDSS photometry is thus an additional source 
of uncertainty, but a small fraction of the total error budget.

	Table \ref{tab:hvs} summarizes the measured and derived properties for the 
16 stars.  We adopt parameters from published echelle spectra when available; 
otherwise we adopt parameters derived from our MMT spectra.  The stellar atmosphere 
parameters for the four bound velocity outliers are new.

\begin{deluxetable*}{cccccccccccc}       % TABLE 1:  HVS Table
\tabletypesize{\scriptsize}
% \rotate
\tablewidth{0pt}
\tablecaption{OBSERVED AND DERIVED HVS PROPERTIES\label{tab:hvs}}
\tablecolumns{12}
\tablehead{
        \colhead{ID} & \colhead{RA, Dec} & 
	\colhead{$v_{\rm helio}$} & \colhead{\teff } &
        \colhead{\logg } & \colhead{\vsini } & \colhead{$M_g$} &
        \colhead{$g_0$} & \colhead{$d_{\rm helio}$} &
        \colhead{$\mu_{\rm RA}, \mu_{\rm Dec}$} & \colhead{$p_{\rm GC}$} &
        \colhead{Ref} \\
         & (J2000) & (\kms) & (K) & (cm s$^{-2}$) & (\kms) & (mag) & (mag) 
	 & (kpc) & (\mas) & &
        }
        \startdata
HVS 1  &  9:07:44.99, $+$02:45:06.9 & $831.1\pm 5.7$ & $11192\pm 450$ & $3.91\pm0.20$ & $158\pm36$ & $-0.36\pm0.31$ & $19.69\pm0.023$ & $102\pm15 $ & $+0.08\pm0.26$, $-0.12\pm0.22$ & 0.47 & 1,13 \\
HVS 2  &  9:33:20.87, $+$44:17:05.5 & $708.0\pm15.0$ & $44561\pm 675$ & $5.23\pm0.12$ &  \nodata   & $+2.22\pm0.30$ & $18.56\pm0.013$ & $ 19\pm2.6$ & $-7.33\pm0.58$, $+2.28\pm0.55$ & 0.00 & 2    \\
HVS 3  &  4:38:12.77, $-$54:33:11.9 & $723.0\pm 3.0$ & $23000\pm1000$ & $3.95\pm0.10$ & $ 55\pm 2$ & $-2.57\pm0.30$ & $16.36\pm0.20 $ & $ 61\pm10 $ & $+0.52\pm0.58$, $+1.65\pm0.57$ & 0.07 & 3,8,9\\
HVS 4  &  9:13:01.00, $+$30:51:19.9 & $600.9\pm 6.2$ & $14547\pm 598$ & $4.15\pm0.21$ & $ 77\pm40$ & $-0.71\pm0.33$ & $18.34\pm0.023$ & $ 64\pm9.8$ & $-0.23\pm0.36$, $-0.42\pm0.36$ & 0.71 & 4,13 \\
HVS 5  &  9:17:59.47, $+$67:22:38.3 & $545.5\pm 4.3$ & $12000\pm 350$ & $3.89\pm0.13$ & $133\pm 7$ & $-0.67\pm0.25$ & $17.58\pm0.032$ & $ 45\pm5.2$ & $+0.55\pm0.61$, $-0.44\pm0.59$ & 0.16 & 4,12 \\
HVS 6  & 11:05:57.45, $+$09:34:39.4 & $609.4\pm 6.8$ & $12190\pm 537$ & $4.30\pm0.23$ & $170\pm55$ & $+0.25\pm0.27$ & $18.94\pm0.024$ & $ 55\pm6.9$ & $+0.05\pm0.57$, $+0.31\pm0.97$ & 0.33 & 5,13 \\
HVS 7  & 11:33:12.13, $+$01:08:24.8 & $526.9\pm 3.0$ & $12000\pm 500$ & $3.80\pm0.10$ & $ 55\pm 2$ & $-0.95\pm0.26$ & $17.63\pm0.015$ & $ 52\pm6.4$ & $+1.00\pm0.82$, $-0.55\pm1.04$ & 0.19 & 5,10 \\
HVS 8  &  9:42:14.03, $+$20:03:22.0 & $499.3\pm 2.9$ & $11000\pm1000$ & $3.75\pm0.25$ & $320\pm60$ & $-0.69\pm0.40$ & $17.93\pm0.016$ & $ 53\pm9.8$ & $-0.82\pm1.16$, $-0.04\pm0.49$ & 0.39 & 6,11 \\
HVS 9  & 10:21:37.08, $-$00:52:34.7 & $616.8\pm 5.1$ & $11637\pm 520$ & $3.84\pm0.21$ & $306\pm72$ & $-0.71\pm0.34$ & $18.64\pm0.023$ & $ 74\pm12 $ & $-1.26\pm0.74$, $-0.25\pm0.70$ & 0.95 & 6,13 \\
HVS 10 & 12:03:37.85, $+$18:02:50.3 & $467.9\pm 5.6$ & $11278\pm 524$ & $4.38\pm0.23$ & $ 37\pm60$ & $+0.65\pm0.24$ & $19.24\pm0.024$ & $ 52\pm5.8$ & $-1.07\pm0.36$, $-0.58\pm0.42$ & 0.86 & 6,13 \\
HVS 12 & 10:50:09.60, $+$03:15:50.6 & $552.2\pm 6.6$ & $12098\pm 622$ & $4.62\pm0.28$ & $ 78\pm88$ & $+0.55\pm0.28$ & $19.63\pm0.024$ & $ 66\pm8.5$ & $-0.40\pm0.36$, $+0.31\pm0.34$ & 0.08 & 7,13 \\
HVS 13 & 10:52:48.31, $-$00:01:33.9 & $569.3\pm 6.1$ & $11054\pm 775$ & $4.00\pm0.35$ & $238\pm43$ & $-0.10\pm0.40$ & $20.01\pm0.021$ & $105\pm19 $ & $-0.90\pm0.38$, $+0.46\pm0.44$ & 0.03 & 7,13 \\
B 434  & 11:02:24.37, $+$02:50:02.8 & $443.9\pm 2.9$ & $10232\pm 382$ & $3.93\pm0.20$ & $117\pm42$ & $+0.06\pm0.27$ & $18.00\pm0.016$ & $ 39\pm4.8$ & $-1.61\pm0.28$, $-0.26\pm0.42$ & 0.08 & 6    \\
B 485  & 10:10:18.82, $+$30:20:28.1 & $408.1\pm 4.8$ & $16167\pm 542$ & $4.02\pm0.12$ & $ 88\pm69$ & $-1.36\pm0.30$ & $16.06\pm0.016$ & $ 30\pm4.3$ & $-1.66\pm0.52$, $-1.15\pm0.38$ & 0.50 & 6    \\
B 711  & 14:20:01.94, $+$12:44:04.7 & $273.7\pm 5.4$ & $11004\pm 298$ & $4.27\pm0.15$ & $ 60\pm76$ & $+0.72\pm0.26$ & $16.92\pm0.016$ & $ 17\pm2.0$ & $-0.96\pm0.80$, $+1.55\pm0.86$ & 0.00 & 6    \\
B 733  & 14:49:55.58, $+$31:03:51.3 & $348.8\pm 2.3$ & $10522\pm 301$ & $4.25\pm0.14$ & $240\pm28$ & $+0.87\pm0.24$ & $15.67\pm0.020$ & $  9\pm1.0$ & $-1.77\pm0.79$, $-3.71\pm0.89$ & 0.00 & 6    \\
        \enddata
\tablerefs{ (1) \citet{brown05}; (2) \citet{hirsch05}; (3) \citet{edelmann05}; (4) \citet{brown06}; (5) \citet{brown06b};
(6) \citet{brown07b}; (7) \citet{brown09a}; (8) \citet{przybilla08}; (9) \citet{bonanos08}; (10) \citet{przybilla08b};
(11) \citet{lopezmorales08}; (12) \citet{brown12c}; (13) \citet{brown14} }
 \end{deluxetable*}

\begin{deluxetable}{ccccrc}       % TABLE 2:  OBSERVATIONS
\tabletypesize{\scriptsize}
\tablewidth{0pt}
\tablecaption{LIST OF {\it HST} OBSERVATIONS\label{tab:obs}}
\tablecolumns{6}
\tablehead{
        \colhead{ID} & \colhead{UT Date} & \colhead{Instr} & \colhead{Filter} & \colhead{P.A.} & \colhead{Exptime } \\ 
         & & & & (deg) & (s)
        }
        \startdata
  HVS 1 & 2006-10-11 &  ACS &   F814W & $ -80.2$ & $522\times5$ \\
\nodata & 2009-10-05 &  ACS & \nodata & $ -82.9$ & $540\times5$ \\
\nodata & 2013-02-21 & WFC3 & \nodata & $ -80.2$ & $612\times8$ \\
  HVS 2 & 2006-10-04 &  ACS &   F814W & $ -56.2$ & $532\times4$ \\
\nodata & 2009-11-03 &  ACS & \nodata & $ -56.2$ & $560\times4$ \\
\nodata & 2012-11-22 & WFC3 & \nodata & $ 142.8$ & $627\times8$ \\
  HVS 3 & 2006-07-08 &  ACS &  F850LP & $-145.2$ & $257\times6$ \\
\nodata & 2009-12-23 &  ACS & \nodata & $  35.0$ & $300\times6$ \\
\nodata & 2012-09-01 & WFC3 & \nodata & $ 124.8$ & $388\times12$ \\
  HVS 4 & 2006-11-06 &  ACS &   F814W & $ -82.0$ & $390\times5$ \\
\nodata & 2009-11-07 &  ACS & \nodata & $ -86.0$ & $410\times5$ \\
\nodata & 2013-02-13 & WFC3 & \nodata & $  24.8$ & $618\times8$ \\
  HVS 5 & 2006-09-01 &  ACS &   F814W & $ -26.1$ & $199\times7$ \\
\nodata & 2009-08-30 &  ACS & \nodata & $ -25.9$ & $200\times7$ \\
  HVS 6 & 2009-11-07 & WFC3 &   F606W & $ 159.3$ & $290\times6$ \\
\nodata & 2012-11-19 & WFC3 & \nodata & $ 159.6$ & $466\times5$ \\
  HVS 7 & 2009-12-04 & WFC3 &   F606W & $ 157.5$ & $475\times3, 90\times3$ \\
\nodata & 2012-12-05 & WFC3 & \nodata & $ 157.8$ & $531\times3, 160\times3$ \\
  HVS 8 & 2009-11-17 & WFC3 &   F606W & $ 139.4$ & $460\times3, 120\times3$ \\
\nodata & 2012-11-26 & WFC3 & \nodata & $ 138.8$ & $494\times3, 200\times3$ \\
  HVS 9 & 2009-11-07 & WFC3 &   F606W & $ 151.8$ & $290\times6$ \\
\nodata & 2012-11-25 & WFC3 & \nodata & $ 152.1$ & $366\times6$ \\
 HVS 10 & 2010-02-23 & WFC3 &   F606W & $ 103.4$ & $379\times6$ \\
\nodata & 2013-03-06 & WFC3 & \nodata & $ 103.3$ & $609\times4$ \\
 HVS 12 & 2009-11-08 & WFC3 &   F606W & $ 155.2$ & $627\times4$ \\
\nodata & 2012-11-30 & WFC3 & \nodata & $ 155.2$ & $613\times4$ \\
 HVS 13 & 2009-11-13 & WFC3 &   F606W & $ 154.5$ & $619\times4$ \\
\nodata & 2012-11-29 & WFC3 & \nodata & $ 154.5$ & $613\times4$ \\
  B 434 & 2009-11-06 & WFC3 &   F606W & $ 155.3$ & $580\times3, 120\times3$ \\
\nodata & 2012-11-22 & WFC3 & \nodata & $ 155.6$ & $536\times3, 160\times3$ \\
  B 485 & 2009-10-30 & WFC3 &   F606W & $ 161.4$ & $563\times3, 25\times3$ \\
\nodata & 2012-11-20 & WFC3 & \nodata & $ 147.8$ & $675\times3, 35\times3$ \\
  B 711 & 2009-12-18 & WFC3 &   F606W & $ 170.5$ & $526\times3, 50\times3$ \\
\nodata & 2012-12-16 & WFC3 & \nodata & $ 170.8$ & $608\times3, 75\times3$ \\
  B 733 & 2009-12-12 & WFC3 &   F606W & $-173.6$ & $563\times3, 16\times3$ \\
\nodata & 2012-12-09 & WFC3 & \nodata & $-173.3$ & $653\times3, 30\times3$ \\
        \enddata
 \end{deluxetable}

\section{PROPER MOTION MEASUREMENTS}

\subsection{{\it HST} Observations}

	We imaged the 16 stars using the {\it HST} ACS and WFC3/UVIS instruments
starting in 2006 September and ending in 2013 March.  We obtained images in a few
different ways; Table \ref{tab:obs} lists the full set of observations.

	We first observed \hbox{HVS 1} - \hbox{HVS 5} using ACS in 2006 (proposal ID 
10824).  Each star was allocated one orbit, and was observed with a set of 4 - 7 
dithered exposures.  Exposure times were chosen to maximize counts on background 
galaxies while not saturating the HVSs.  We obtained a second epoch of observations 
with ACS in 2009 (proposal ID 11782).  For \hbox{HVS 1} - \hbox{HVS 4}, we then 
obtained a third epoch of observations using WFC3 (proposal ID 12503).  We 
allocated two orbits of time per star in the third epoch, but dropped \hbox{HVS 5} 
because its field contained few useful background galaxies.  The time baseline of 
observations for \hbox{HVS 1} - \hbox{HVS 4} is thus about 6.2 yrs, and for 
\hbox{HVS 5} it is 3.0 yrs.

	We observed the other 12 stars exclusively with WFC3 beginning in 2009-2010 
(proposal ID 11589).  Each star was allocated one orbit of time and observed with a 
set of dithered exposures.  For the brightest 6 stars, however, we divided the 
exposures into a set of short $\simeq$1 min and deep $\simeq$10 min exposures.  Our 
goal was to optimally expose the bright HVSs and faint background galaxies in the 
short and deep exposures, respectively, and to tie their astrometric frames together 
using the intermediate brightness stars available in both sets of exposures.  In 
practice, the finite number of intermediate brightness stars limits the accuracy of 
this approach.  We obtained a second epoch of observations with WFC3 in 2012-2013 
(proposal ID 12662) in the same way.  The time baseline of the observations is 3.0 
yrs for these 12 stars.

	We made an effort to use the same telescope orientation across all epochs of 
imaging in order to minimize the impact of changes in charge-transfer efficiency 
(CTE) and errors in the distortion solution.  The effects of CTE increase with time 
as on-orbit radiation damage creates more and more charge traps in the CCD silicon 
lattice.  The result is that star and galaxy images are trailed along the CCD 
read-out direction.  If we observe in the same orientation in all epochs, the CTE 
systematic is in the same direction and we minimize its impact on our differential 
position measurements.  Unfortunately, we were unable to use the same guide stars 
for \hbox{HVS 2}, \hbox{HVS 3}, and \hbox{HVS 4} in all epochs due to changes in the 
{\it HST} guide star catalog.  This issue forced a change in orientation, which 
exacerbates the astrometric effects of CTE.  Table \ref{tab:obs} lists the position 
angle (P.A.), defined as increasing East of due North, for each observation.

\subsection{Image Reduction and Analysis}

	We begin our data reduction procedure by downloading flat-fielded images
from the Mikulski Archive for Space Telescopes.  Because the CTE correction for
WFC3/UVIS images and ACS subarrays is not integrated into the standard pipeline, we
run the \citet{anderson10} pixel-based correction by hand.  The CTE correction is
calibrated on hot pixels and their charge trails, and is successful at removing the
trails behind stars, cosmic rays, and hot pixels, and restoring the flux to the
stellar images.

	We start our analysis with the first-epoch images.  We use empirical models 
of the ACS and WFC3/UVIS point spread functions (PSFs) to measure positions for 
stars in the CTE-corrected frames of each first epoch exposure.  We then correct the 
positions for geometric distortion using the model in \citet{anderson06} for ACS and 
the model in \citet{bellini11} for WFC3/UVIS.  Finally, we cross identify the stars 
and define a linear transformation from the distortion-corrected positions of each 
star in each exposure into the distortion-corrected frame of the first exposure.

	Since all of the first-epoch images were taken in a single orbit, we can 
safely use the stars to define the transformations.  This cannot be done for the 
later epochs, since the stars are all moving, and we need to know the motion of each 
target star in an inertial frame.  Therefore, we use the background galaxies to 
define the transformation from each exposure into the reference frame.  Although all 
stars can be fit with the same PSF, each galaxy has its own unique distribution of 
light and must be measured with its own template.  An additional complication is 
that some images were taken at different orientations or with different instruments, 
such that the PSF for a given object might be different from epoch to epoch, 
affecting the observed distribution of light.  For this reason, we construct 
templates for each galaxy with a deconvolved model, such that when we fit the 
template to the pixels in an image, we first convolve it with the PSF appropriate 
for that location in that detector.  This approach ensures that the stars (which are 
simply delta functions when deconvolved) and galaxies are measured consistently 
within an exposure.

	We construct the star and galaxy templates from the exposures in the 
first-epoch data set.  The transformations (based on the stars) enable us to 
accurately map the pixels of all exposures into the reference frame.  For each 
source, we collect all the pixels that map to within the $11\times11$ region about 
each source (galaxies and stars).  We then use iterative forward modeling to 
construct a deconvolved template for each source.  The template for the stars is 
simply a delta function.  The template for the galaxies is a smooth empirical image 
that, when convolved with the PSF, best described the distribution of light in the 
contributing exposures.  We allowed the galaxies to have an additional point source 
at their centers, but most did not need one.

	We then use these templates, convolved with the appropriate PSF, to measure 
consistent positions for every source in every exposure in each epoch.  We examine 
the fitting residuals for each source and reject the measurements that were clearly 
contaminated by cosmic rays or unidentified warm pixels.  Measured positions are in 
the raw frame and must be corrected for geometric distortion, as mentioned above.  
Next, we use the positions for the galaxies to define the linear transformation from 
the distortion-corrected frame of each exposure into that of the first exposure of 
the first epoch.  We examine the transformation residuals for each galaxy and reject 
those galaxies with unreliable and inconsistent positions.  This leaves us with 10 
to 15 high quality galaxies for each field.  From the residuals, we infer that the 
transformations are typically good to $\pm0.93$ mas, which corresponds to a 0.31 
\mas\ systematic uncertainty in the reference frame for a 3 yr time baseline.  The 
systematic uncertainty is a noise floor to all of our measurements.

	Finally, we use these galaxy-based transformations to map the position for 
each star in each exposure into the reference frame, and then solve for the proper 
motions.  Mathematically, the proper motion calculation boils down to a linear least 
squares fit to the x and y positions vs.\ time of observation, which we know 
exactly.  We convert the x and y pixel motions into \mas\ using the appropriate 
camera pixel scale, image rotation angle, and time baseline of observations. Proper 
motion uncertainties are determined from the scatter in x and y positions at each 
epoch, and added in quadrature to the uncertainty of the galaxy reference frame.  
Total measurement uncertainties range from $\pm0.35$ \mas\ for the case of \hbox{HVS 
1}, which has a 6.36 yr time baseline of observations, to $\pm1.33$ \mas\ for the 
case of \hbox{HVS 7}, which has only 8 useable reference galaxies.  Our final proper 
motion measurements are in Table \ref{tab:hvs}.

\subsection{Comparison with Previous Measurements}

	Our proper motion measurements are a 6-fold improvement over previous 
measurements.  One of the most accurate proper motion catalogs available today is 
the UCAC4, which covers the entire sky to a depth of $R=16$ mag \citep{zacharias13}.  
One of our stars, B~733, is bright enough to have a UCAC4 measurement.  The UCAC4 
total proper motion for B~733, $\mu_{\rm UCAC4,B~733} = 13.9 \pm 6.6$ \mas, agrees 
to within 1.5$\sigma$ of our measurement $\mu_{\rm B~733} = 4.1 \pm 1.2$ \mas.  
Another catalog is PPMXL \citep{roeser10}, which combines 2MASS and USNO-B 
astrometry.  Only B~733 is bright enough to have a 2MASS measurement, and the PPMXL 
total proper motion $\mu_{\rm PPMXL,B~733} = 6.3 \pm 6.8$ \mas\ is in perfect 
agreement with our measurement.
	The SDSS survey probes much deeper and, when combined with USNO-B astrometry 
\citep{monet03}, provides proper motion measurements for all of our sample except 
\hbox{HVS 3}.  The average SDSS-USNO-B proper motion uncertainty for our stars is 
$\pm5.3$ \mas, and only one star, \hbox{HVS 5}, has a SDSS proper motion that 
differs from zero at greater than 1.5$\sigma$ significance.  In contrast, our {\it 
HST\/} measurements have an average proper motion uncertainty of $\pm0.8$ \mas, and 
thus are 6.6 times more accurate.

\begin{figure*}		% FIGURE 3:  Proper Motion
 \plotone{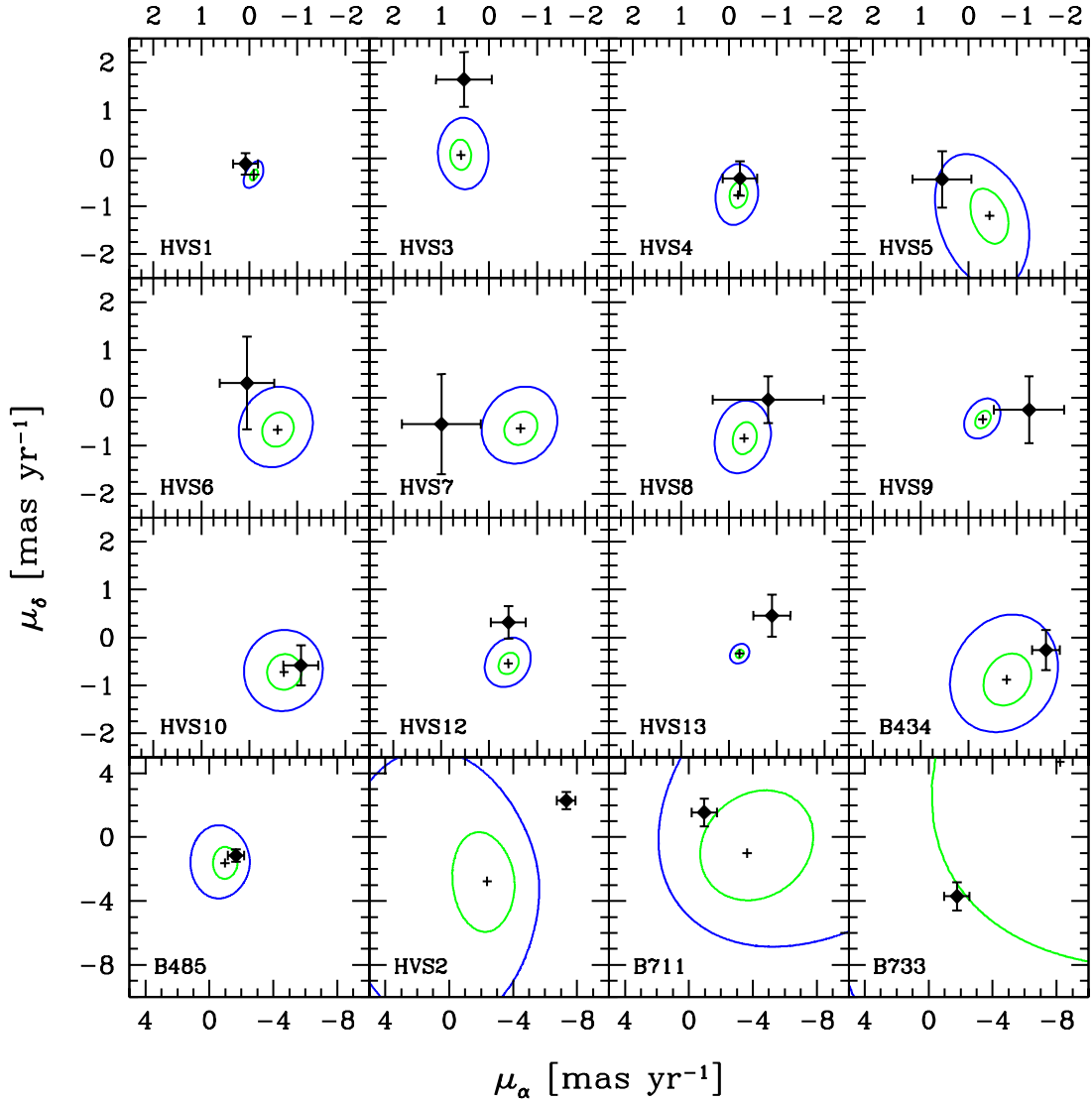}
 \caption{ \label{fig:pmall}
	Proper motions in context:  assuming a fixed distance, radial velocity, 
right ascension, and declination for each star (those listed in Table 
\ref{tab:hvs}), we compare the measured proper motions (points with errorbars) to 
the locus of trajectories that pass within 8 kpc (green ellipse) and 20 kpc (blue 
ellipse) of the Galactic center.  Trajectories that pass through the Galactic center 
are marked by a +.  All calculations use the \citet{kenyon14} potential model.  
To evaluate the likelihood of origin requires that we account for all of the 
measurement errors, as seen in Figure \ref{fig:xyall}. }
 \end{figure*}

\subsection{Proper Motions in Context}

	At first glance, the stars have small proper motions and thus largely radial 
trajectories.  Twelve of the stars have proper motions statistically consistent with 
zero.  Only \hbox{HVS 2}, B~434, B~485, and B~733 have proper motions that differ 
from zero at better than 3$\sigma$ significance.

	For distant stars, however, the reflex motion caused by the Sun orbiting the 
Milky Way can dominate the apparent proper motion.  The direction and amplitude of 
solar reflex motion depends on the location and distance of the star.  Thus, a 
correct interpretation of the proper motion measurements requires that we calculate 
trajectories.

	As a first step, we adopt a fixed distance, radial velocity, right 
ascension, and declination (those listed in Table \ref{tab:hvs}) for each star, and 
then calculate each trajectory backward in time for all possible proper motions that 
cross the Galactic plane.  We use the \citet{kenyon14} three component 
bulge-disk-halo potential model for the trajectory calculations.  This potential 
model uniquely fits observed mass measurements from the Galactic center to the outer 
halo, but was originally constructed for a 220 \kms\ circular velocity 
\citep{kenyon08}.  Updating the disk mass $M_d=6\times10^{10}$ \msun\ and disk 
radial scale length $a_d=2.75$ kpc yields a flat rotation curve of 235 \kms\ 
consistent with the most recent circular velocity measurements \citep{reid14}.  We 
refer to this potential model as the \citet{kenyon14} model, and use it in all of 
our trajectory calculations.

	Assuming a fixed distance, radial velocity, right ascension, and declination 
(those listed in Table \ref{tab:hvs}), Figure \ref{fig:pmall} compares our measured 
proper motions to the results of our trajectory calculations.  The green and blue 
ellipses show the locus of proper motions with trajectories that pass within 8 and 
20 kpc, respectively, of the Galactic center.  Statistically, all of our stars have 
proper motions consistent with a Milky Way origin.  Our strongest constraints are 
for the three closest stars that have significantly non-zero proper motions:  
\hbox{HVS 2}, B~711, and B~733.  To evaluate the likelihood of a more exact origin 
requires that we account for all of the measurement errors, in proper motion, radial 
velocity, and distance.

\section{ORIGIN OF HYPERVELOCITY STARS}

	With proper motions in hand, we can now address the question of origin.  We 
expect that the fastest unbound stars, as discussed above, are most likely HVSs 
ejected by the MBH in the Galactic center.  Lower velocity stars can be explained by 
alternative scenarios, such as runaway ejections from the disk.  Importantly, 
runaways with unbound speeds are most likely to be those stars ejected in the 
direction of Galactic rotation from the outer disk \citep{heber08, bromley09, 
kenyon14}.  Thus we can distinguish between the runaway and MBH ejection scenarios 
if we know where the trajectories of our stars cross the Galactic plane.

	The stars in our sample were selected on the basis of their extreme radial 
velocities.  Their trajectories in the Galactic frame will therefore point 
away from the Sun unless they have proper motions that significantly deviate from 
the Solar reflex motion.

	In reality, the errors in proper motion are too large to pinpoint the 
location of ejection with the desired accuracy.  To put the measurement 
uncertainties in context, consider a star's tangential velocity $v_{\rm tan} = 4.74 
d \mu$, where $d$ is the heliocentric distance in kpc and $\mu$ is the proper motion 
in \mas.  The typical star in our sample has a median distance of 50 kpc and a 
proper motion error of $\pm$0.8 \mas, thus an error in tangential velocity of 
$\pm$190 \kms.  For the median total space velocity of 650 \kms, the uncertainty in 
the angle of trajectory is $\pm16\arcdeg$.  Over a distance of 50 kpc, the 
uncertainty in position is thus $\pm$15 kpc.

	While we would like to test whether an observed trajectory goes exactly 
through the Galactic center, the errors preclude locating a trajectory to any one 
point.  Instead, we use the measurements to try to disprove a Galactic center 
origin.  Our approach is to calculate the statistical consistency of the 
measurements with a Galactic center trajectory.  We obtain the same answer to this 
question whether we calculate trajectories from the star backward in time or from 
the Galactic center forward in time, because the proper motion and distance errors 
are the same either way.  Measurement errors make either the Galactic center or the 
star look very blurry.

	We use a Monte Carlo calculation to account for all of the measurement 
uncertainties, and visualize the results in a distribution of Galactic 
plane-crossing locations.  For each star, we draw 1,000,000 current velocities and 
distances assuming that the measured proper motion, radial velocity, and distance 
have Gaussian random uncertainties.  We then calculate each trajectory backward in 
time and record where it crosses the Galactic plane.

\begin{figure*}		% FIGURE 4:  Galactic Plane Crossing
 \plotone{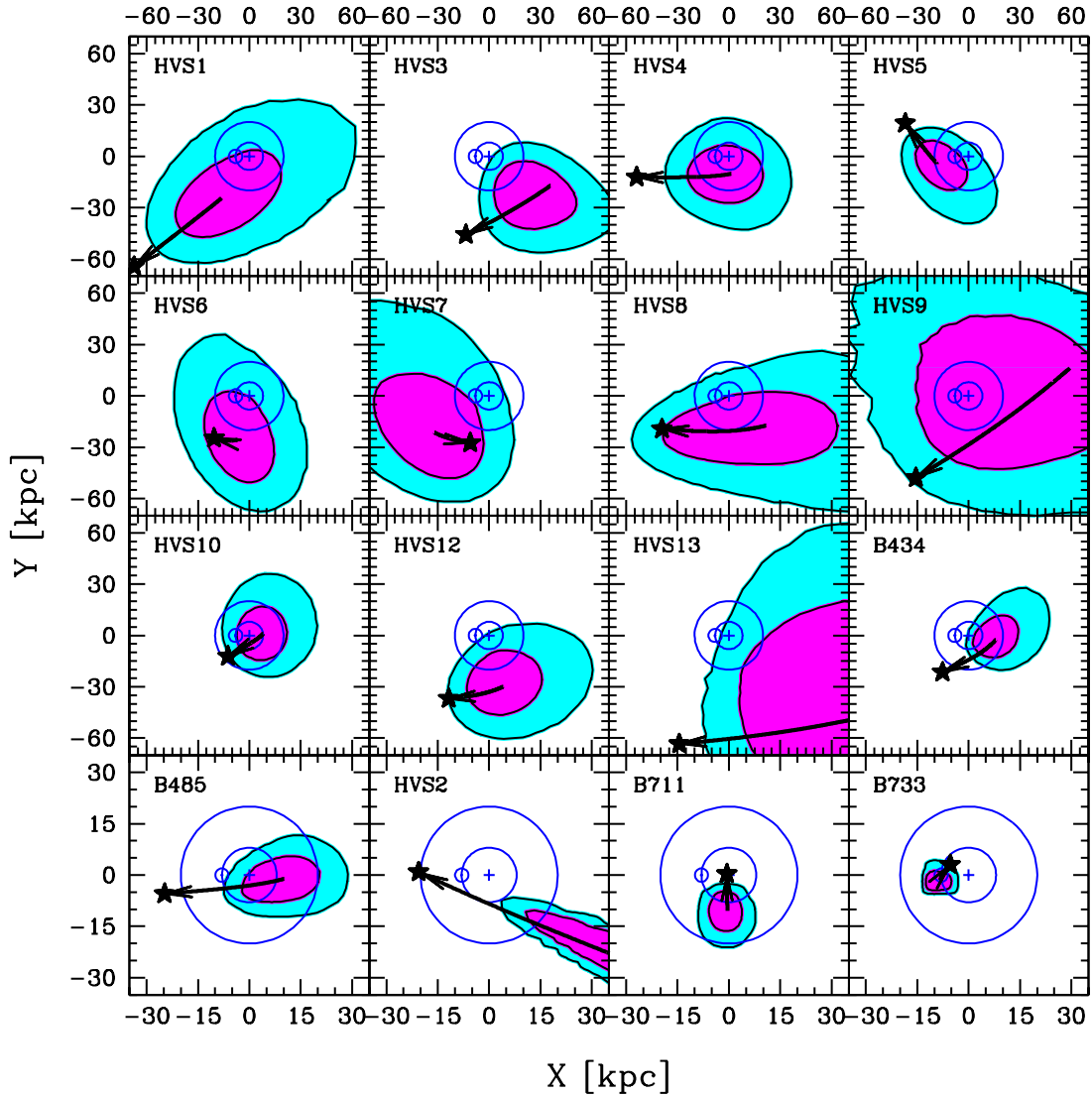}
 \caption{ \label{fig:xyall}
	Distribution of Galactic plane-crossing locations for each star, calculated 
assuming Gaussian random uncertainties in the measured proper motions, radial 
velocities, and distances.  For context, the large blue circles show the $R=8$ solar 
circle and $R=20$ kpc outer edge of the disk.  The Sun is at $X=-8$ kpc (small blue 
circle), the Galactic center is marked with +, and the disk rotates clockwise in 
this plot.  Black stars show the present location of each star, and the solid black 
lines with arrows show their trajectories.  Magenta and cyan ellipses are the 1- and 
2-$\sigma$ likelihood regions where the stars cross the Galactic plane. }
 \end{figure*}

	Figure \ref{fig:xyall} presents the resulting distribution of Galactic 
plane-crossing locations for our sample.  A black star shows the present location of 
each star, and a solid black line shows its trajectory.  Magenta and cyan ellipses 
show the 1$\sigma$ and 2$\sigma$ likelihood plane-crossing regions, respectively.  
We determine these regions by calculating the density of plane-crossing 
locations in bins of X and Y, and solving for the density level thresholds that 
contain 68.26\% and 95.44\% of all the crossings within them.  The contours are 
centered on the mode of the plane-crossing distributions, which can differ from the 
mean trajectory (black line).  Drawing both a large distance and a large proper 
motion can result in a very large tangential motion and thus an extreme 
plane-crossing location.  We discard those trajectories that fail to cross the 
Galactic plane within the main sequence lifetime of each star.  For reference, a 3 
\msun\ star has a 365 Myr main sequence lifetime in the Padova tracks; for the 
special cases HVS~2 and HVS~3, we allow flight times of 1 Gyr and 100 Myr, 
respectively.  These flight time constraints exclude fewer than 1\% of trajectories 
for most of the sample.

	Because distance and proper motion are the most uncertain parameters, the
size and shape of the Galactic plane-crossing likelihood regions are driven by the
distance and proper motion uncertainties.  There is no visible change in the
Galactic plane-crossing likelihood regions if we use a different potential model,
specifically the one based on \citet{gnedin05} and updated in \citet{gnedin14}.  
Similarly, a triaxial halo potential model, like that expected from cosmological
simulations, changes the trajectories very little.

	Half of our stars have trajectories that formally cross the Galactic plane 
at $R>20$ kpc, outside of the Milky Way disk, but all of our stars include a part of 
the Milky Way within their 1$\sigma$ plane-crossing likelihood regions.  Only one 
star, \hbox{HVS 7}, has a trajectory moving opposite to the direction of Galactic 
rotation. \hbox{HVS 7} is also our most uncertain proper motion measurement.  
Because a counter-rotation trajectory requires an unphysically large Galactic plane 
ejection velocity, we consider the astrometric link between the short- and 
long-exposures of \hbox{HVS 7} suspect.  Many more stars, including \hbox{HVS 1}, 
\hbox{HVS 4}, \hbox{HVS 10}, and B~485, have very radial trajectories.

	We reject the Galactic center origin hypothesis if the Galactic center 
trajectory falls outside the 3$\sigma$ ($p_{GC}<0.0026$) threshold of the 
distribution (Figure \ref{fig:xyall}).  The value of $p_{GC}$ is determined as the 
fraction of all possible orbits with the density of plane-crossings below the value 
at the GC bin (X=0,Y=0).  This respresents the consistency of the Galactic center 
trajectory with the measurement uncertainties.  By this estimate of probability, 13 
of our stars are consistent with a Galactic center origin within about the $2\sigma$ 
($p_{GC}\ge0.046$) confidence level.  As a sanity check, we obtain essentially the 
same result if we ignore the potential model and simply compare a perfectly radial 
trajectory against the observed tangential velocities.  Tangential velocity error 
drives the probability.  Only the stars \hbox{HVS 2}, B~711, and B~733 have 
measurements inconsistent with a Galactic center origin at $>$3$\sigma$ 
($p_{GC}<0.0026$) confidence:  they are runaways.  We discuss these objects in more 
detail below.

\section{INDIVIDUAL OBJECTS}

\subsection{HVS 2}

	HVS~2 is a helium rich sdO star with a $708\pm15$ \kms\ heliocentric 
radial velocity \citep{hirsch05}.  The two proposed explanations for its extreme 
motion are an ejection by the central MBH \citep{hirsch05, perets09a} or a Type Ia 
supernova explosion \citep{justham09, wang09}.

	HVS~2 has a significant proper motion of $(\mu_{\rm RA}, \mu_{\rm Dec}) = 
(-7.33\pm0.58, 2.28\pm 0.55)$ \mas.  Our measurement is based on three epochs of 
imaging and therefore has a very well understood error distribution.  Given 
\hbox{HVS 2's} $18.5\pm2.6$ kpc heliocentric distance \citep{hirsch05}, the observed 
proper motion corresponds to a $673\pm118$ \kms\ tangential motion.  \hbox{HVS 2's} 
tangential motion is quite similar to its radial velocity.  Thus its space motion in 
the rest frame of the Milky Way is almost exactly 1,000 \kms(!).

	The direction of \hbox{HVS 2's} proper motion takes its trajectory across 
the Galactic plane at $R=76.5$ kpc.  This result implies that \hbox{HVS 2} 
originates from the stellar halo, possibly launched by a Type Ia supernova explosion 
as proposed by \citet{justham09} and \citet{wang09}.  If \hbox{HVS 2's} distance and 
proper motion are 1$\sigma$ smaller, however, its tangential motion is less extreme 
and it crosses the Galactic plane around $R=20$ kpc (see Figure \ref{fig:xyall}).  
Thus an ejection from the disk is a viable possibility.

	A Galactic center origin for \hbox{HVS 2}, on the other hand, is ruled out 
at greater than 3$\sigma$ confidence.  \hbox{HVS 2's} closest approach to the 
Galactic center, given the measurement errors, is $R=4.5$ kpc among our 1,000,000 
Monte Carlo trajectory calculations.  In other words, the extreme velocity of 
\hbox{HVS 2} cannot be explained by a dynamical interaction via gravitational 
interaction with the central MBH.  Instead, it is an example of a hyper-runaway 
star.

	\citet{wang09} perform ejection calculations for helium star companions to 
white dwarfs that explode in Type Ia supernovae.  In this model, the helium stars 
receive a kick perpendicular to their orbital velocity for total ejection velocities 
of 500 -- 650 \kms.  Similarly, \citet{geier13} calculate a 600 \kms\ velocity 
from their Type Ia supernova ejection model.  Given these ejection velocities, the 
Type Ia supernova model cannot explain \hbox{HVS 2's} 1,000 \kms\ motion with a disk 
ejection; \hbox{HVS 2's} trajectory is not even in the direction of Galactic 
rotation (Figure \ref{fig:xyall}).  \citet{tauris15}'s aysmmetric core-collapse 
supernova model can produce 1,000 \kms\ velocities for low mass stars. However, 
there would not be enough time for a low mass star to evolve into an sdO star before 
its core-collapse companion launches it out of the Galaxy.  The remaining 
possibility is that \hbox{HVS 2} was launched by a Type Ia supernova from a halo 
binary traveling at $\sim$400 \kms\ in its current direction of motion.

	Alternative origins seem less likely.  For example, \citet{abadi09} propose
that the tidal disruption of a dwarf galaxy can explain unbound stars.  This
mechanism requires the close peri-center passage of a fairly massive $>10^{10}$
\msun\ dwarf to produce unbound stars \citep{piffl11}.  The non-Galactic-center
trajectory of \hbox{HVS 2}, plus the absence of other unbound stars around it, would
appear to rule out the dwarf galaxy tidal debris origin.
	Additional evidence for a Type Ia supernova explosion, perhaps found in the
abundance pattern of \hbox{HVS 2's} stellar atmosphere, would better support the
supernova origin picture.

\subsection{B~711 and B~733}

	B~711 and B~733 are both bound B-type stars at modest 9 -- 17 kpc distances, 
and thus plausible candidates for being runaway B stars ejected from the disk.  
Alternatively, they could be failed HVSs on bound trajectories from the Galactic 
center.  Their radial velocities in the Galactic frame are 290 and 440 \kms, 
respectively. In the absence of a positive \vsini\ measurement for B~711, it is also 
possible that B~711 could be an evolved low mass star (i.e.\ a hot horizontal branch 
star) orbiting in the stellar halo.  The fact that B~733 is a rapidly rotating 2.5 
\msun\ main-sequence star with a 349 \kms\ heliocentric radial velocity, however, 
requires that it was ejected from a location in the Galaxy with recent star 
formation.

	We measure proper motions that point to a disk runaway origin for both 
stars.  The stars formally cross the Galactic plane at $R=10$ kpc (Figure 
\ref{fig:xyall}).  A Galactic center origin is ruled out at greater than 3$\sigma$ 
confidence for both stars.

	The trajectories of B~711 and B~733 cross the Galactic plane at angles 
nearly perpendicular to Galactic rotation, however, implying disk ejection 
velocities of 533 \kms\ and 441 \kms.  The supernova ejection mechanism is able to 
achieve 500 \kms\ velocities for main-sequence stars only in extreme scenarios 
\citep{tauris15}.  The dynamical ejection mechanism can achieve 500 \kms\ velocities 
for main-sequence stars but requires 3-body interactions with contact binaries 
containing 100 \msun\ stars \citep{gvaramadze09, gvaramadze11}.  Stars with 100 
\msun\ are rare and short-lived. Simulations suggest that perhaps 0.1\% of dynamical 
ejections reach 500 \kms\ velocities \citep{perets12}.  The upshot is that we expect 
to find more Galactic center ejections than disk runaway ejections at these speeds 
\citep{bromley09, perets12, kenyon14}.  The fact that B~711 and B~733 are 
$\simeq$500 \kms\ disk runaway ejections is thus quite intriguing.

	Because extreme runaway ejections require massive stars with relatively 
short lifetimes, we expect B~711 and B~733 to have flight times similar to their 
stellar ages.  Our proper motions correspond to trajectories with 20 - 40 Myr flight 
times from the disk.  Our spectroscopic \logg\ measurements favor young ages for 
both B~711 and B~733 (see Figure \ref{fig:teff}), however the uncertainties are 
large and no statistically meaningful constraint is currently possible.  High 
resolution echelle spectroscopy of B~711 and B~733 would thus be very interesting.  
B~733 is clearly a main-sequence star on the basis of its rapid rotation, and its 
trajectory is thus evidence for an extreme stellar dynamical ejection from the disk 
like that seen for \hbox{HD 271791} \citep{heber08} and \hbox{HIP 60350}
\citep{irrgang10}.

\begin{figure}		% FIGURE 5:  HVS3
 \plotone{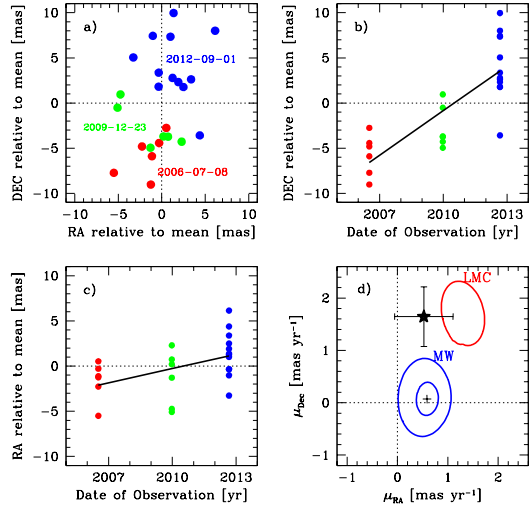}
 \caption{ \label{fig:hvs3}
	a) \hbox{HVS 3} position measurements plotted relative to the mean; we label 
epoch 1 red, epoch 2 green, and epoch 3 blue.  b) Dec positions versus date of 
observation; solid line is the linear least squares fit to the data.  c) Same as b) 
but for RA.  d) Proper motion (star with errorbars) compared to the locus of 
\hbox{HVS 3} proper motions with trajectories that pass with 3 kpc of the LMC (red 
ellipse) and within 8 and 20 kpc of the Milky Way (blue ellipses).  The ellipses are 
calculated for a fixed \hbox{HVS 3} distance; accounting for the distance error, 
\hbox{HVS 3} is consistent with both LMC and Milky Way origins at the 1$\sigma$ 
level.}

 \end{figure}

\subsection{HVS 3}

	HVS~3 is the unbound 9 \msun\ main-sequence B star near the LMC on the sky 
\citep{edelmann05}.  If \hbox{HVS 3} comes from the LMC, then its speed is evidence 
for a MBH hidden somewhere in the LMC \citep{gualandris07}.  If \hbox{HVS 3} comes 
from the Milky Way, then its speed and stellar nature are evidence of a former 
binary HVS ejection \citep{lu07, perets09a}.  Both origins are unlikely in terms of 
ejection rates.

	In 2010, we published a proper motion for \hbox{HVS 3} that pointed to a 
Milky Way origin \citep{brown10b}.  This measurement was based on two epochs of ACS 
imaging with a 3.46 yr time baseline.  The readout direction of the ACS CCD was 
unfortunately aligned with the Milky Way-LMC direction, however, and added an 
additional systematic uncertainty.  We now have a third epoch of WFC3 imaging, 
obtained at an orientation angle 90 deg from the previous data sets, that doubles 
our time baseline.  Our expectation is that \hbox{HVS 3's} intrinsic motion now 
dominates the errors.  We re-process and re-analyze the epoch 1 and 2 data in the 
same way as the epoch 3 data, so that everything is in a common reference frame.

	Figure \ref{fig:hvs3} presents the results.  We plot in panel a) the 
position of \hbox{HVS 3}, relative to its mean position, as measured in each 
individual image.  Different epochs are identified by color, and the scatter in 
positions reflects the underlying precision of our measurements.  We observe 
that \hbox{HVS 3} moves $10.6\pm5.0$ mas, or 0.27 WFC3 pixels, in 6.15 years.  This 
motion is quantified in Figure panels b) and c), which plot the RA and Dec positions 
of \hbox{HVS 3} versus time.  The solid lines in each panel show the linear least 
squares fit to the measurements.  The proper motion of \hbox{HVS 3} is $(\mu_{\rm 
RA}, \mu_{\rm Dec}) = (0.52\pm0.58, 1.65\pm 0.57)$ \mas.

	As fate would have it, our \hbox{HVS 3} proper motion corresponds to a 
physical trajectory that passes in-between the Milky Way and LMC.  Figure 
\ref{fig:hvs3} panel d) is identical to Figure \ref{fig:pmall} except that we now 
draw the locus of \hbox{HVS 3} proper motions with trajectories that pass within 3 
kpc of the LMC (red ellipse).  A 3 kpc radius encompasses the full extent of the LMC 
bar and all of the young clusters proposed by \citet{gualandris07} for the origin of 
\hbox{HVS 3}.  We determine \hbox{HVS 3's} distance to the LMC in Figure 
\ref{fig:hvs3} assuming that the LMC has a mass of $2\times10^{10}$ \msun\ moving on 
an orbit that reproduces the line-of-sight velocity from \citet{vandermarel01} and 
the proper motion from \citet{kallivayalil06} in our Milky Way potential model.  If 
we account for the uncertainties in \hbox{HVS 3} and LMC distances, then both LMC 
and Milky Way origins are consistent with the measurements at the 1$\sigma$ level.

	The ambiguity on origin is driven by our epoch 3 measurements, which exhibit 
a larger shift, and scatter, in position than seen in epochs 1 and 2.  This result 
is visually apparent in Figure \ref{fig:hvs3}.  Fitting only the epoch 1 and 2 
measurements yields $(\mu_{\rm RA}, \mu_{\rm Dec}) = (0.10, 0.89)$ \mas, in 
1$\sigma$ agreement with our previously published value \citep{brown10b}.  Fitting 
only epoch 2 and 3 measurements, on the other hand, yields $(\mu_{\rm RA}, \mu_{\rm 
Dec}) = (1.01, 2.52)$ \mas.  Clearly, the third epoch drives the proper motion to 
larger values, and thus causes \hbox{HVS 3's} trajectory to move away from the Milky 
Way in physical space.  

	We repeat our measurement and analysis for HVS~3 using only the 5 best 
galaxies and obtain essentially the same result.  Our reference frame thus appears 
robust.  We speculate that the F850LP filter, chosen to maximize S/N on the 
background galaxies while not saturating the bright blue star, may be partly to 
blame.  The F850LP filter has a less well-calibrated distortion solution than the 
F814W and F606W filters, which were used for all of our other observations.  The 
other HVSs with three epochs of data all have less scatter between epochs:  the 
proper motions derived from epochs 1 and 2 and derived from epochs 2 and 3 agree at 
the $1\sigma$ level, except for the Dec motion of HVS~2 that differs by $2\sigma$. 
Residual CTE systematic error is also a possible problem for the HVS~3 observations. 
Answering the question of \hbox{HVS 3's} origin will ultimately require a proper 
motion measurement with a longer time baseline of observations, or else a better 
instrument like {\it Gaia}.

\section{CONCLUSION}

	We present {\it HST} proper motion measurements for 16 stars with extreme 
radial velocities, 12 of which are unbound to the Milky Way.  On the basis of 
spectroscopic stellar atmosphere fits, our sample consists of 15 main-sequence B 
stars and 1 helium-rich sdO star located at 10 -- 100 kpc distances in the stellar 
halo. We expect that the fastest stars are likely HVSs ejected by the central MBH, 
and that the lower velocity stars may be runaways ejected from the Milky Way disk. 

	We process our images using the best geometric distortion solutions, CTE 
corrections, and empirical point spread function fits.  Our final proper motions 
have an average uncertainty of $\pm0.8$ \mas, a 6-fold improvement over previous 
measurements.  Twelve of our stars have proper motions consistent with zero, and 
thus largely radial trajectories.

	Given the uncertainties in proper motion and distance, the data allow 
for a wide range of origin locations.  We calculate the statistical consistency of 
the data with a Galactic center trajectory in an attempt to disprove the Galactic 
center origin hypothesis.  We find that a Galactic center trajectory remains 
consistent with the measurements for 13 of our stars within the 2$\sigma$ confidence 
level.  Only the stars \hbox{HVS 2}, B~711, and B~733 are inconsistent with a 
Galactic center origin at $>$3$\sigma$ confidence, and thus runaways.

	HVS~2 is an unbound sdO star whose trajectory points from the stellar halo, 
possibly explained by a Type Ia supernova explosion as proposed by \citet{justham09} 
and \citet{wang09}.  Its 1,000 \kms\ motion is in some tension with the supernova 
ejection model, but can be explained if \hbox{HVS 2} was ejected from a halo orbit.  
B~711 and B~733, on the other hand, are B stars with trajectories that clearly point 
from the stellar disk.  These two stars are thus runaway stars, and their 
trajectories provide strong evidence for $\sim$500 \kms\ ejections from the disk.

	Our third epoch of imaging for \hbox{HVS 3}, the unbound 9 \msun\ B star 
near the LMC, yields a larger proper motion than previously measured 
\citep{brown10b} and thus a trajectory further from the Milky Way.  Accounting for 
the uncertainty in \hbox{HVS 3's} distance, its trajectory is now equally consistent 
with a Milky Way and LMC origin.

	In the future, better constraints on HVS trajectories and origin will come 
from improved proper motion measurements.  Doubling the amount of spectra will not 
significantly improve distance estimates, for example, but doubling the time 
baseline of imaging will in principle double the precision of the proper motions.  
Even better, in 2017 {\it Gaia} will begin releasing proper motions for all of the 
HVSs.  {\it Gaia}'s predicted precision ranges from $\pm0.03$ \mas\ for 16th mag 
stars like \hbox{HVS 3} and B~733, to $\pm0.3$ \mas\ for a 20th mag star like 
\hbox{HVS 1}, and with improved accuracy compared to our measurements.  {\it Gaia} 
will thus provide interesting constraints for all of the known HVSs.

	If the unbound HVSs are indeed ejected from the Galactic center, we can use
their trajectories to probe the shape and orientation of the Milky Way's dark matter
halo \citep{gnedin05}. HVSs are effective test particles that traverse the Galaxy to
$\sim$100 kpc distances. If the Galactic potential is triaxial, as predicted by cold
dark matter simulations, the present motion of HVSs must deviate from being
precisely radial.  With a sufficient number of HVSs in different directions on the
sky, and proper motions accurate to better than 0.1 \mas, it may be possible to
measure the two axis ratios and three direction angles of the triaxial halo.

{\it Facilities:} \facility{HST (ACS, WFC3)}; \facility{MMT (Blue Channel 
Spectrograph)}

\acknowledgements

	We thank the referee for constructive scientific comments.  This research
makes use of SAO/NASA's Astrophysics Data System Bibliographic Services.  This
research makes use of the Sloan Digital Sky Survey, which is funded by the Alfred
P.\ Sloan Foundation, the Participating Institutions, and the U.S.\ Dept.\ of
Energy.  H.E.B.\ acknowledges support by NASA through grants GO-11589 and GO-12503
from the Space Telescope Science Institute, which is operated by AURA, Inc., under
NASA contract NAS~5-26555.  This work was supported in part by the Smithsonian
Institution.

	% REFERENCES
% \clearpage
% \bibliographystyle{/home/wbrown/lib/apj} \bibliography{/home/wbrown/text/Ref}

\begin{thebibliography}{63}
\expandafter\ifx\csname natexlab\endcsname\relax\def\natexlab#1{#1}\fi

\bibitem[{{Abadi} {et~al.}(2009){Abadi}, {Navarro}, \& {Steinmetz}}]{abadi09}
{Abadi}, M.~G., {Navarro}, J.~F., \& {Steinmetz}, M. 2009, \apjl, 691, L63

\bibitem[{{Abt} {et~al.}(2002){Abt}, {Levato}, \& {Grosso}}]{abt02}
{Abt}, H.~A., {Levato}, H., \& {Grosso}, M. 2002, \apj, 573, 359

\bibitem[{{Ahn} {et~al.}(2014){Ahn}, {Alexandroff}, {Allende Prieto},
  {et~al.}}]{ahn14}
{Ahn}, C.~P., {Alexandroff}, R., {Allende Prieto}, C., {et~al.} 2014, \apjs,
  211, 17

\bibitem[{{Anderson} \& {Bedin}(2010)}]{anderson10}
{Anderson}, J. \& {Bedin}, L.~R. 2010, \pasp, 122, 1035

\bibitem[{{Anderson} \& {King}(2006)}]{anderson06}
{Anderson}, J. \& {King}, I.~R. 2006, {``PSFs, Photometry, and Astronomy for
  the ACS/WFC''}, Tech. rep., STScI

\bibitem[{{Behr}(2003)}]{behr03b}
{Behr}, B.~B. 2003, \apjs, 149, 101

\bibitem[{{Bellini} {et~al.}(2011){Bellini}, {Anderson}, \&
  {Bedin}}]{bellini11}
{Bellini}, A., {Anderson}, J., \& {Bedin}, L.~R. 2011, \pasp, 123, 622

\bibitem[{{Bellini} {et~al.}(2014){Bellini}, {Anderson}, {van der Marel},
  {et~al.}}]{bellini14}
{Bellini}, A., {Anderson}, J., {van der Marel}, R.~P., {et~al.} 2014, \apj,
  797, 115

\bibitem[{{Blaauw}(1961)}]{blaauw61}
{Blaauw}, A. 1961, \bain, 15, 265

\bibitem[{{Bonanos} {et~al.}(2008){Bonanos}, {L{\'o}pez-Morales}, {Hunter}, \&
  {Ryans}}]{bonanos08}
{Bonanos}, A.~Z., {L{\'o}pez-Morales}, M., {Hunter}, I., \& {Ryans}, R.~S.~I.
  2008, \apjl, 675, L77

\bibitem[{{Bressan} {et~al.}(2012){Bressan}, {Marigo}, {Girardi},
  {et~al.}}]{bressan12}
{Bressan}, A., {Marigo}, P., {Girardi}, L., {et~al.} 2012, \mnras, 427, 127

\bibitem[{{Bromley} {et~al.}(2009){Bromley}, {Brown}, {Geller}, \&
  {Kenyon}}]{bromley09}
{Bromley}, B.~C., {Brown}, W.~R., {Geller}, M.~J., \& {Kenyon}, S.~J. 2009,
  \apj, 706, 925

\bibitem[{{Brown} {et~al.}(2010){Brown}, {Anderson}, {Gnedin},
  {et~al.}}]{brown10b}
{Brown}, W.~R., {Anderson}, J., {Gnedin}, O.~Y., {et~al.} 2010, \apjl, 719, L23

\bibitem[{{Brown} {et~al.}(2012{\natexlab{a}}){Brown}, {Cohen}, {Geller}, \&
  {Kenyon}}]{brown12c}
{Brown}, W.~R., {Cohen}, J.~G., {Geller}, M.~J., \& {Kenyon}, S.~J.
  2012{\natexlab{a}}, \apjl, 754, L2

\bibitem[{{Brown} {et~al.}(2013){Brown}, {Cohen}, {Geller}, \&
  {Kenyon}}]{brown13b}
---. 2013, \apj, 775, 32

\bibitem[{{Brown} {et~al.}(2009){Brown}, {Geller}, \& {Kenyon}}]{brown09a}
{Brown}, W.~R., {Geller}, M.~J., \& {Kenyon}, S.~J. 2009, \apj, 690, 1639

\bibitem[{{Brown} {et~al.}(2012{\natexlab{b}}){Brown}, {Geller}, \&
  {Kenyon}}]{brown12b}
---. 2012{\natexlab{b}}, \apj, 751, 55

\bibitem[{{Brown} {et~al.}(2014){Brown}, {Geller}, \& {Kenyon}}]{brown14}
---. 2014, \apj, 787, 89

\bibitem[{{Brown} {et~al.}(2005){Brown}, {Geller}, {Kenyon}, \&
  {Kurtz}}]{brown05}
{Brown}, W.~R., {Geller}, M.~J., {Kenyon}, S.~J., \& {Kurtz}, M.~J. 2005,
  \apjl, 622, L33

\bibitem[{{Brown} {et~al.}(2006{\natexlab{a}}){Brown}, {Geller}, {Kenyon}, \&
  {Kurtz}}]{brown06}
---. 2006{\natexlab{a}}, \apjl, 640, L35

\bibitem[{{Brown} {et~al.}(2006{\natexlab{b}}){Brown}, {Geller}, {Kenyon}, \&
  {Kurtz}}]{brown06b}
---. 2006{\natexlab{b}}, \apj, 647, 303

\bibitem[{{Brown} {et~al.}(2007{\natexlab{a}}){Brown}, {Geller}, {Kenyon},
  {Kurtz}, \& {Bromley}}]{brown07a}
{Brown}, W.~R., {Geller}, M.~J., {Kenyon}, S.~J., {Kurtz}, M.~J., \& {Bromley},
  B.~C. 2007{\natexlab{a}}, \apj, 660, 311

\bibitem[{{Brown} {et~al.}(2007{\natexlab{b}}){Brown}, {Geller}, {Kenyon},
  {Kurtz}, \& {Bromley}}]{brown07b}
---. 2007{\natexlab{b}}, \apj, 671, 1708

\bibitem[{{Edelmann} {et~al.}(2005){Edelmann}, {Napiwotzki}, {Heber},
  {Christlieb}, \& {Reimers}}]{edelmann05}
{Edelmann}, H., {Napiwotzki}, R., {Heber}, U., {Christlieb}, N., \& {Reimers},
  D. 2005, \apjl, 634, L181

\bibitem[{{Geier} {et~al.}(2013){Geier}, {Marsh}, {Wang}, {et~al.}}]{geier13}
{Geier}, S., {Marsh}, T.~R., {Wang}, B., {et~al.} 2013, \aap, 554, A54

\bibitem[{{Girardi} {et~al.}(2004){Girardi}, {Grebel}, {Odenkirchen}, \&
  {Chiosi}}]{girardi04}
{Girardi}, L., {Grebel}, E.~K., {Odenkirchen}, M., \& {Chiosi}, C. 2004, \aap,
  422, 205

\bibitem[{{Gnedin} {et~al.}(2005){Gnedin}, {Gould}, {Miralda-Escud{\'e}}, \&
  {Zentner}}]{gnedin05}
{Gnedin}, O.~Y., {Gould}, A., {Miralda-Escud{\'e}}, J., \& {Zentner}, A.~R.
  2005, \apj, 634, 344

\bibitem[{{Gnedin} {et~al.}(2014){Gnedin}, {Ostriker}, \&
  {Tremaine}}]{gnedin14}
{Gnedin}, O.~Y., {Ostriker}, J.~P., \& {Tremaine}, S. 2014, \apj, 785, 71

\bibitem[{{Gualandris} \& {Portegies Zwart}(2007)}]{gualandris07}
{Gualandris}, A. \& {Portegies Zwart}, S. 2007, \mnras, 376, L29

\bibitem[{{Gvaramadze} \& {Gualandris}(2011)}]{gvaramadze11}
{Gvaramadze}, V.~V. \& {Gualandris}, A. 2011, \mnras, 410, 304

\bibitem[{{Gvaramadze} {et~al.}(2009){Gvaramadze}, {Gualandris}, \& {Portegies
  Zwart}}]{gvaramadze09}
{Gvaramadze}, V.~V., {Gualandris}, A., \& {Portegies Zwart}, S. 2009, \mnras,
  396, 570

\bibitem[{{Heber}(2009)}]{heber09}
{Heber}, U. 2009, \araa, 47, 211

\bibitem[{{Heber} {et~al.}(2008){Heber}, {Edelmann}, {Napiwotzki}, {Altmann},
  \& {Scholz}}]{heber08}
{Heber}, U., {Edelmann}, H., {Napiwotzki}, R., {Altmann}, M., \& {Scholz},
  R.-D. 2008, \aap, 483, L21

\bibitem[{{Hills}(1988)}]{hills88}
{Hills}, J.~G. 1988, \nat, 331, 687

\bibitem[{{Hirsch} {et~al.}(2005){Hirsch}, {Heber}, {O'Toole}, \&
  {Bresolin}}]{hirsch05}
{Hirsch}, H.~A., {Heber}, U., {O'Toole}, S.~J., \& {Bresolin}, F. 2005, \aap,
  444, L61

\bibitem[{{Huang} \& {Gies}(2006)}]{huang06a}
{Huang}, W. \& {Gies}, D.~R. 2006, \apj, 648, 580

\bibitem[{{Irrgang} {et~al.}(2010){Irrgang}, {Przybilla}, {Heber}, {Nieva}, \&
  {Schuh}}]{irrgang10}
{Irrgang}, A., {Przybilla}, N., {Heber}, U., {Nieva}, M.~F., \& {Schuh}, S.
  2010, \apj, 711, 138

\bibitem[{{Irrgang} {et~al.}(2013){Irrgang}, {Wilcox}, {Tucker}, \&
  {Schiefelbein}}]{irrgang13}
{Irrgang}, A., {Wilcox}, B., {Tucker}, E., \& {Schiefelbein}, L. 2013, \aap,
  549, A137

\bibitem[{{Justham} {et~al.}(2009){Justham}, {Wolf}, {Podsiadlowski}, \&
  {Han}}]{justham09}
{Justham}, S., {Wolf}, C., {Podsiadlowski}, P., \& {Han}, Z. 2009, \aap, 493,
  1081

\bibitem[{{Kallivayalil} {et~al.}(2006){Kallivayalil}, {van der Marel},
  {Alcock}, {et~al.}}]{kallivayalil06}
{Kallivayalil}, N., {van der Marel}, R.~P., {Alcock}, C., {et~al.} 2006, \apj,
  638, 772

\bibitem[{{Kenyon} {et~al.}(2014){Kenyon}, {Bromley}, {Brown}, \&
  {Geller}}]{kenyon14}
{Kenyon}, S.~J., {Bromley}, B.~C., {Brown}, W.~R., \& {Geller}, M.~J. 2014,
  \apj, 793, 122

\bibitem[{{Kenyon} {et~al.}(2008){Kenyon}, {Bromley}, {Geller}, \&
  {Brown}}]{kenyon08}
{Kenyon}, S.~J., {Bromley}, B.~C., {Geller}, M.~J., \& {Brown}, W.~R. 2008,
  \apj, 680, 312

\bibitem[{{Kurtz} \& {Mink}(1998)}]{kurtz98}
{Kurtz}, M.~J. \& {Mink}, D.~J. 1998, \pasp, 110, 934

\bibitem[{{Leonard}(1991)}]{leonard91}
{Leonard}, P.~J.~T. 1991, \aj, 101, 562

\bibitem[{{L{\'o}pez-Morales} \& {Bonanos}(2008)}]{lopezmorales08}
{L{\'o}pez-Morales}, M. \& {Bonanos}, A.~Z. 2008, \apjl, 685, L47

\bibitem[{{Lu} {et~al.}(2007){Lu}, {Yu}, \& {Lin}}]{lu07}
{Lu}, Y., {Yu}, Q., \& {Lin}, D.~N.~C. 2007, \apjl, 666, L89

\bibitem[{{Marigo} {et~al.}(2008){Marigo}, {Girardi}, {Bressan}, {Groenewegen},
  {Silva}, \& {Granato}}]{marigo08}
{Marigo}, P., {Girardi}, L., {Bressan}, A., {Groenewegen}, M.~A.~T., {Silva},
  L., \& {Granato}, G.~L. 2008, \aap, 482, 883

\bibitem[{{Monet} {et~al.}(2003){Monet}, {Levine}, {Canzian},
  {et~al.}}]{monet03}
{Monet}, D.~G., {Levine}, S.~E., {Canzian}, B., {et~al.} 2003, \aj, 125, 984

\bibitem[{{Perets}(2009)}]{perets09a}
{Perets}, H.~B. 2009, \apj, 698, 1330

\bibitem[{{Perets} \& {Subr}(2012)}]{perets12}
{Perets}, H.~B. \& {Subr}, L. 2012, \apj, 751, 133

\bibitem[{{Piffl} {et~al.}(2014){Piffl}, {Scannapieco}, {Binney},
  {et~al.}}]{piffl14}
{Piffl}, T., {Scannapieco}, C., {Binney}, J., {et~al.} 2014, \aap, 562, A91

\bibitem[{{Piffl} {et~al.}(2011){Piffl}, {Williams}, \& {Steinmetz}}]{piffl11}
{Piffl}, T., {Williams}, M., \& {Steinmetz}, M. 2011, \aap, 535, A70

\bibitem[{{Portegies Zwart}(2000)}]{portegies00}
{Portegies Zwart}, S.~F. 2000, \apj, 544, 437

\bibitem[{{Poveda} {et~al.}(1967){Poveda}, {Ruiz}, \& {Allen}}]{poveda67}
{Poveda}, A., {Ruiz}, J., \& {Allen}, C. 1967, Bol.\ Obs\ Tonantzintla
  Tacubaya, 4, 860

\bibitem[{{Przybilla} {et~al.}(2008{\natexlab{a}}){Przybilla}, {Nieva},
  {Heber}, {et~al.}}]{przybilla08}
{Przybilla}, N., {Nieva}, M.~F., {Heber}, U., {et~al.} 2008{\natexlab{a}},
  \aap, 480, L37

\bibitem[{{Przybilla} {et~al.}(2008{\natexlab{b}}){Przybilla}, {Nieva},
  {Tillich}, {et~al.}}]{przybilla08b}
{Przybilla}, N., {Nieva}, M.~F., {Tillich}, A., {et~al.} 2008{\natexlab{b}},
  \aap, 488, L51

\bibitem[{{Reid} {et~al.}(2014){Reid}, {Menten}, {Brunthaler},
  {et~al.}}]{reid14}
{Reid}, M.~J., {Menten}, K.~M., {Brunthaler}, A., {et~al.} 2014, \apj, 783, 130

\bibitem[{{Roeser} {et~al.}(2010){Roeser}, {Demleitner}, \&
  {Schilbach}}]{roeser10}
{Roeser}, S., {Demleitner}, M., \& {Schilbach}, E. 2010, \aj, 139, 2440

\bibitem[{{Tauris}(2015)}]{tauris15}
{Tauris}, T.~M. 2015, \mnras, 448, L6

\bibitem[{{van der Marel}(2001)}]{vandermarel01}
{van der Marel}, R.~P. 2001, \aj, 122, 1827

\bibitem[{{Wang} \& {Han}(2009)}]{wang09}
{Wang}, B. \& {Han}, Z. 2009, \aap, 508, L27

\bibitem[{{Zacharias} {et~al.}(2013){Zacharias}, {Finch}, {Girard},
  {et~al.}}]{zacharias13}
{Zacharias}, N., {Finch}, C.~T., {Girard}, T.~M., {et~al.} 2013, \aj, 145, 44

\bibitem[{{Zheng} {et~al.}(2014){Zheng}, {Carlin}, {Beers}, {et~al.}}]{zheng14}
{Zheng}, Z., {Carlin}, J.~L., {Beers}, T.~C., {et~al.} 2014, \apjl, 785, L23

\end{thebibliography}

\end{document}